\documentstyle[PASJadd,epsfig]{PASJ95} 

\newcommand{\vol}{52}
\newcommand{\no}{3}
\newcommand{\lett}{}
\newcommand{\spage}{}
\newcommand{\rdate}{1999 December 22}
\newcommand{\adate}{2000 February 17}

\begin{document}
\setcounter{page}{1}

\title{Slim Disk Model for Soft X-Ray Excess and Variability\break
of Narrow-Line Seyfert 1 Galaxies}

\author{Shin {\sc Mineshige}, Toshihiro {\sc Kawaguchi}
\\
{\it Department of Astronomy, Graduate School of Science,
Kyoto University, Sakyo-ku, Kyoto 606-8502}
\\
{\it  E-mail(SM): minesige@kusastro.kyoto-u.ac.jp}
\\
Mitsuru {\sc Takeuchi}
\\
{\it Astronomical Institute, Osaka-Kyoiku University, 
Asahigaoka, Kashiwara 582-8582}
\\
\centerline{and}
\\
Kiyoshi {\sc Hayashida}
\\
{\it Department of Earth and Space Sciences, Graduate School of Science,
Osaka University, Toyonaka 560-0043}}

\abst{
Narrow-line Seyfert 1 galaxies (NLS1s) exhibit
extreme soft X-ray excess and large variability.
We argue that
both features can be basically accounted for by the slim disk model.
We assume that a central black-hole mass in NLS1 is relatively small,
$M \sim 10^{5-7}M_\odot$, and that a disk shines nearly at
the Eddington luminosity, $L_{\rm E}$.
Then, the disk becomes a slim disk and exhibits
the following distinctive signatures:
(1) The disk luminosity (particularly of X-rays)
is insensitive to mass-flow rates, $\dot M$,
since the generated energy is partly
carried away to the black hole by trapped photons in accretion flow.
(2) The spectra are multi-color blackbody.
The maximum blackbody temperature is
$T_{\rm bb} \simeq 0.2(M/10^5 M_\odot)^{-1/4}$ keV,
and the size of the blackbody emitting region is small,
$r_{\rm bb} \lsim 3 r_{\rm S}$
(with $r_{\rm S}$ being Schwarzschild radius)
even for a Schwarzschild black hole.
(3) All the ASCA observation data of NLS1s fall onto the region
of $\dot M/(L_{\rm E}/c^2)>10$ 
(with $L_{\rm E}$ being the Eddington luminosity)
on the ($r_{\rm bb},T_{\rm bb}$) plane,
supporting our view that a slim disk emits soft X-rays at 
$\sim L_{\rm E}$ in NLS1s.
(4) Magnetic energy can be amplified, at most, up to the
equipartition value with the trapped radiation energy
which greatly exceeds radiation energy emitted from the disk.
Hence, energy release by consecutive magnetic reconnection
will give rise to substantial variability in soft X-ray emission.
}

\kword{accretion, accretion disks --- black holes --- 
galaxies: active --- galaxies: Seyfert}

\maketitle

\section{Introduction}

The terminology of
{\it narrow-line} Seyfert 1 galaxies (NLS1s) originates from
their relatively $narrow$ optical Balmer-line emission
with $v_{\rm BLR} \lsim 2000$km s$^{-1}$ 
compared with those of usual Seyfert 1 galaxies with
broad-line emission, BLS1s
(see a concise review by Brandt 1999 and references therein).  
Their optical line ratios and remarkable Fe{\sc ii} lines
(Osterbrock, Pogge 1985; Halpern, Oke 1987; 
Grupe et al. 1999) are also distinct from those of BLS1s.
NLS1s exhibit unique X-ray properties ($e.g.$, Grupe et al. 1998);
they are characterized by large soft X-ray excess 
(Pounds et al. 1996; Otani et al. 1996; Leighly 1999b)
and a good correlation is known to exist between
the strength of soft excess and FWHM of optical Balmer lines
(Boller et al. 1996; Laor et al. 1997).  
Rapid soft/hard X-ray variability is another signature characterizing 
NLS1s (Otani et al. 1996; Boller et al. 1997; Leighly 1999a).
NLS1s are not rare but comprise a significant part (say, 20\%)
of the Seyfert 1 galaxies and it seems that a group of NLS1s 
is smoothly connected to the class of BLS1s (Brandt, Boller 1998).

It is often suggested that
the strength of the soft X-ray excess relative to the hard X-ray power
law appears to be directly related to the `primary eigenvector'
of Boroson and Green (1992), which represents the strongest set of
optical emission line correlations ($e.g.$, Brandt 1999).
In this sense, NLS1s are located at the extreme end of the primary eigenvector.  The UV line properties of NLS1s also 
fits this scheme (Wills et al. 1999).  
Then, what is the control parameter which drives
the primary eigenvector?
What physical factor distinguishes NLS1s from BLS1s?

There are several models proposed to account for
the narrowness of otherwise broad-line emission 
(see Boller et al. 1996 for extensive discussion).
Successful models should account for a smooth continuation
between NLS1s and BLS1s.
Among them, one fascinating and probably most promising explanation is
that NLS1s contain
relatively less massive black holes (with $M \sim 10^{5-7}M_\odot$).
Moderate brightness of NLS1s, $L \sim 10^{43-45}$erg s$^{-1}$,
implies similar mass-flow rates, $\dot M$, to those of BLS1s.
It then follows that the ratio of $L/L_{\rm E}$ (or $\dot M/M$)
is relatively large for NLS1s.
The control parameter which drives the eigen vector could be 
the fraction of the Eddington rate at which 
the supermassive black hole is accreting, ${\dot M}/(L_{\rm E}/c^2)$.  

For small black-hole masses, narrow Balmer line emission of NLS1s
can be understood, provided that the broad-line clouds 
(which emit broad-line emission in BLS1s)
are bound in the potential by a central black hole (Laor et al. 1997).
Since the radius of the broad-line clouds are roughly scaled as
$r_{\rm BLR} \propto L^{0.5}$ from the reverberation mapping
(Kaspi et al. 1996),
the circular velocity of broad-line clouds, 
$\sim \sqrt{GM/r_{\rm BLR}}$, is systematically smaller 
for smaller $M$ at a constant $L$ (and thus constant $r_{\rm BLR}$),
yielding narrow Balmer emission.

For a luminosity close to the Eddington,
\begin{equation}
   L_{\rm E} \simeq 1.2\times 10^{43} M_5~{\rm erg~s}^{-1},
\end{equation}
where $M_5 \equiv M/(10^5M_\odot)$,
what do we then expect theoretically for the disk structure and its spectra?
It is known that for such a high luminosity
advective energy transport dominates over
radiative cooling (Abramowicz et al. 1988;
section 10.3 of Kato et al. 1998).
Such a disk is named as the slim disk, since it is moderately
geometrically thick.  

The slim disk is an optically thick version of ADAF
(advection-dominated accretion flow)
and should not be confused with
optically thin ADAF proposed for low-luminosity AGNs
(see section 10.2 of Kato et al. 1998 and references therein).
Unlike the optically thin ADAF
the observable features of the slim disk have been poorly investigated.
Szuszkiewicz et al. (1996) was the first to start investigation
along this line (see also Wang et al. 1999), 
and we give more detailed discussion 
on the observational consequences in the present study.
Watarai et al. (2000a) studied the super-critical accretion flow 
in the context of Galactic black-hole candidates (GBHC),
showing some unique observational features of the slim disk.

We focus on the AGN case in the present study.
Methods of calculations are given in section 2.
We then discuss the non-standard emission properties of the slim disk
in the context of NLS1s in section 3.
We next turn to the subject of large variability and discuss 
its origin in relation to magnetic-field activity in section 4.
Outstanding issues will be discussed in section 5.
The final section is devoted to conclusions.
Throughout the present study we use the normalization
of $\dot M$
with $\dot M_{\rm crit} \equiv L_{\rm E}/c^2$; i.e.,
\begin{eqnarray}
  {\dot m} \equiv {\dot M}/{\dot M}_{\rm crit}
       &\!\!\sim\!\!& {\dot M}/(1.3\times 10^{22} M_5~ {\rm g~s}^{-1})   \cr
       &\!\!\sim\!\!& {\dot M}/(2.5\times 10^{-4}M_5M_\odot{\rm ~yr}^{-1}).
\end{eqnarray}

\section{Basic equations and methods of calculations}

We first solve the steady-state, transonic disk structure.
The methods are the same as those adopted in Matsumoto et al. (1984)
and Watarai et al. (2000a), 
but we briefly repeat the description below.

For the purpose of the present study, 
it is essential to solve the full-set equations,
since the emission from the innermost region is in question.
(Approximate solutions, such as the self-similar solutions,
do not properly treat the inner boundary conditions nor
regularity conditions.)
The transonic nature of the accretion flow should also be
carefully treated.
We do not include any general relativistic (GR) effects
except for adopting the pseudo-Newtonian potential
by Paczy\'nski and Witta (1980),  $\psi = -GM/(r-r_{\rm S})$.
For discussion of GR effects, see section 3.5.
We use cylindrical coordinates, ($r,\varphi,z$).

Throughout this paper, we used vertically integrated 
(or height-averaged) equations.
In fact, we numerically confirmed that $H/r < 1$ for $\dot m \lsim 100$,
where $H$ is the scale-height of the disk.
Certainly, $H/r > 1$ for even higher mass accretion rate, but for such
high $\dot m$,
photon diffusion time over $H$ becomes longer than accretion timescale 
so that the disk height should be significantly reduced (see section 5.1). 
The height-average approach may not be very precise
and we eventually need fully 2-D models.
The same problem arises in optically thin ADAF, 
for which Narayan \& Yi (1995) found that 
the 2-D solutions of the exact non-height-integrated equations agree quite 
well with those of the simplified height-integrated equations, 
provided that `height-integration' is done along $\theta$ at constant spherical
radius, rather than along $z$ at constant cylindrical radius. The
height-integrated equations therefore are a fairly accurate
representation of quasi-spherical ADAFs.

We construct the vertically integrated
equations, using integrated variables, such as 
surface density, $\Sigma \equiv \int\rho dz$, and integrated pressure,
$\Pi \equiv \int pdz$.
Vertically integrated mass, momentum, and angular-momentum
conservations lead
\begin{equation}
\label{mass}
   -2\pi r\Sigma v_r={\dot M}= {\rm const.},    
\end{equation}
\begin{equation}
\label{r-mom}
   v_r{dv_r\over dr}+{1\over \Sigma}{d\Pi\over dr}
   ={\ell^2-\ell_{\rm K}^2\over r^3}
   -{\Pi\over \Sigma}{d{\rm ln}\Omega_{\rm K}\over dr},   
\end{equation}
and
\begin{equation}
\label{ang-mom}
    {\dot M}(\ell-\ell_{\rm in})=-2\pi r^2T_{r\varphi},  
\end{equation}
respectively, where 
$\Omega (=v_\varphi/r)$ and $\Omega_{\rm K}
  [=(GM/r)^{1/2}/(r-r_{\rm S})$]
are the angular frequency of the gas flow
and the Keplerian angular frequency
in the pseudo-Newtonian potential,
$\ell$ ($=rv_\varphi$) is specific angular momentum,
$\ell_{\rm K}$ ($\equiv r^2\Omega_{\rm K}$)
 is the Keplerian angular momentum,
$\ell_{\rm in}$ is the specific angular momentum finally 
swallowed by the black hole,
and
$r_{\rm S}\equiv 2GM/c^2 \simeq 3 \times 10^{10}M_5$ cm
is the Schwarzschild radius.
As to the viscous stress tensor we adopt the prescription
$T_{r\varphi} \equiv -\alpha \Pi$ with the viscosity
parameter, $\alpha = 0.1$.
Note, however, that
the resultant disk structure does not sensitively depend on 
$\alpha$ (Watarai et al. 2000b).

The vertically integrated energy equation is symbolically 
written as
\begin{equation}
\label{energy}
       Q_{\rm adv}^-=Q_{\rm vis}^+-Q_{\rm rad}^-,  
\end{equation}
where $Q_{\rm adv}^- [\propto \Sigma v_rT(ds/dr)$] 
is the advective cooling with $s$ being specific entropy,
and other two terms on the right-hand side represent
viscous heating and radiative cooling, respectively.

The outer boundary conditions are imposed at 
$r = 2.0 \times 10^3 r_{\rm S}$, where
each physical quantity is taken as that of the standard disk.
We confirmed that the disk structure near the outer boundary 
can be very accurately described by the standard-disk relation.
When calculating spectra (see below),
we extrapolate the temperature distribution 
up to $r=2.0\times 10^4 r_{\rm S}$
using the standard-disk relation, $T_{\rm eff} \propto r^{-3/4}$.
The basic equations are integrated by the semi-implicit method
from the outer boundary 
to the inner one taken at $r = 1.01 r_{\rm S}$.  
The flow is subsonic far outside ($r \gg r_{\rm S}$)
but is supersonic just near the black hole.
The solution should satisfy the regularity condition at the
transonic radius, at about 2.7$r_{\rm S}$,
and we adopt the free boundary conditions at the inner edge.

\section{Observational features of a slim disk}

\subsection{Unique temperature profile}

How do structural changes occur from the standard to
the slim disk when $\dot m$ increases?
The solid lines in the upper panels of
figure 1 represent the temperature profiles 
of the disks for various values of $\dot m$
for the black-hole mass of $M=10^5 M_\odot$.
For comparison, we also plot with the dashed lines
the temperatures expected from the standard-disk relation
(Shakura, Sunyaev 1973),
\begin{equation}
\label{SS}
 T_{\rm eff} \simeq 10^{6.5}
          M_5^{-\frac 14}
	   {\dot m}^{\frac 14}
    \left(\frac{r}{r_{\rm S}}\right)^{-\frac 34}
    \left(1-\sqrt{r_{\rm in}\over r}\right)^{\frac 14}{\rm K},
\end{equation}
where $M_5 \equiv M/10^5M_\odot$.

% Fig.1
\begin{figure}[t]
\hbox{\psfig{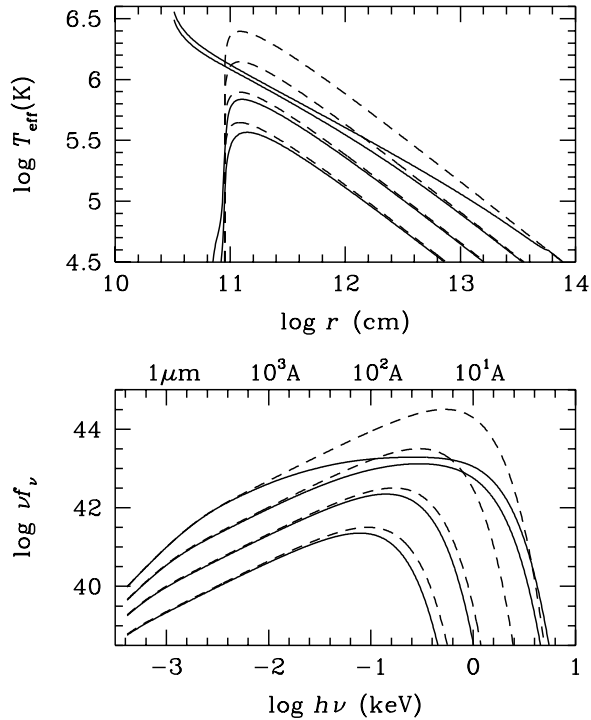}}
\caption
{
The temperature profiles (upper) and the emergent 
spectra (lower) of the calculated disks with different
mass-flow rate; $\dot m=1.0$, 10.0, 100.0, and 1000.0
from the below.
As $\dot m$ increases, regions with a flat temperature profile 
(extending down to $\log r\sim 10.5$ or $r\sim r_{\rm S}$)
expands from the innermost region
and the disk spectra become flatter at the peak.
For comparison, we plot the temperature profiles expected by
the standard-disk relation [equation (\ref{SS})]
and their emergent spectra by the dashed lines.
}
\end{figure}

We note three characteristics:
\begin{enumerate}
\item
In contrast with the standard disk having an effective
temperature profile of $T_{\rm eff} \propto r^{-3/4}$,
the slim disk exhibits flatter slopes,
$T_{\rm eff} \propto r^{-1/2}$ (Watarai, Fukue 1999).
As $\dot m$ increases, the part with a flatter temperature 
profile expands outward; i.e., up to 
$r\sim 100 r_{\rm S}$ for $\dot m\sim 100$, and
$r\sim 1000 r_{\rm S}$ for $\dot m\sim 1000$.

\item When the innermost parts become advection-dominated
at $\dot m \gsim 50$, substantial amount of radiation 
is expected from inside the marginally stable last stable orbit, 
$r_{\rm ms} = 3r_{\rm S} \sim 10^{11}$cm.
This can be understood as follows:
For $\dot m \ll 1$, the flow is very subsonic far outside,
while it should become supersonic near the black hole.
A sudden jump in the radial accretion velocity across
the transonic radius (at $\sim 2.7r_{\rm S}$) results in 
an abrupt density drop inside the transonic radius
(note $\dot M \propto v_r\Sigma$ is kept constant).
The innermost region become of low density, 
thus producing little emission (emissivity goes like $\rho^2$).
For $\dot m \gsim 50$, conversely,
the flow velocity outside the transonic radius
is already comparable to the sound speed,
thus the velocity and density jump across the transonic radius being small.
Therefore, huge, continuous mass supply makes it possible to keep
the innermost region (at $r < 3r_{\rm S}$) optically thick
and substantial amount of radiation can be produced there as a result.

\item Consequently, 
the inner disk temperature (which makes a hotter peak in the spectrum)
gets systematically higher than the values expected 
by the standard relation.  The size of the region emitting
with the maximum energy becomes small, however.
\end{enumerate}

\subsection{Spectral features}

Once temperature distributions are obtained,
we can calculate emergent spectra by simply summing up 
a contribution from each tiny portion of the disk.
Such scattering dominated, hot disks as those considered here
do not always emit blackbody radiation (Czerny, Elvis 1987;
Beloborodov 1998).  
However, we simply
assume blackbody radiation to see the effects of transonic flow
(discussed later).

The results are plotted in the lower panels of figure 1.
As $\dot m$ increases,
the spectra, $\nu S_\nu$, get flatter and flatter around the peak,
$\nu S_\nu \sim \nu^0$.
(Still, $\nu S_\nu \sim \nu^{4/3}$ in the optical bands.)
According to Grupe et al. (1998) soft X-ray selected AGNs
(including a number of NLS1s) tend to exhibit bluer optical/UV 
(around 1400 \AA) spectra 
when compared with those of hard X-ray selected ones
and the maximum slope is $S_\nu \propto \nu^{0.3}$.
This is consistent with our calculations.

The most significant problem lies in $a_{\rm os}$, or the optical
to soft X-ray flux ratio (say 3000A to 0.2 keV). 
High $L/L_{\rm E}$ models predict a very flat $a_{\rm os}$, --1 or flatter, 
as shown in figure 1, whereas the observed slope is more like --1.5 
(see Table 1 of Brandt et al. 2000, which includes many NLS1s).
Thus, the observed soft X-ray emission is a factor of 10, or more, 
weaker than high $L/L_{\rm E}$ disk models predict. 
In fact, this problem already exists for normal AGNs
(Laor et al. 1997) and is actually more severe for NLS1s because
of the flatter predicted $a_{\rm os}$.
This does not necessarily rule out the disk model, but
it does suggest a significantly reduced efficiency of the emission from the
inner parts of the accretion disk (e.g. Laor 1998).

NLS1s have a steeper soft (0.1--2 keV) X-ray slope,
but since they have a pretty normal $a_{\rm os}$, 
this slope is actually mostly because the hard X-ray component is weak 
and not because the soft component is strong (Laor et al. 1997).
The slim disk model shows that
exponential roll-over extends over 1keV,
thus producing large power-law indices in the ROSAT band.  
The inclusion of Compton scattering within
the disk will increase the photon indices further 
(Ross et al. 1992) which will be discussed in section 3.5.
Fairly strong correlation between the optical and 
soft X-ray emission is naturally explained since both are from
the same optically thick disk.

If we include a hard power-law component in the calculation, we
expect a larger hard power law index in the ASCA bands 
for a larger soft excess as a result of more efficient cooling 
of electrons via inverse Compton scattering.
Since the hard power-law slope is determined by the Compton $y$-parameter
which is related to the energy amplifications of soft photons
(e.g. Rybicki, Lightman 1979), 
less (or more) energy input to high-energy electrons 
compared with that to soft photons leads to
a larger (smaller) $y$ and thus
a larger (smaller) photon index in the hard X-ray bands; 
$\alpha_{\rm x} \gsim 2$ 
($\alpha_{\rm x}< 2$) as is observed in the soft (hard) state 
of GBHC (Mineshige et al. 1995).
Systematically large $\alpha_{\rm x}$'s in NLS1s in the ASCA bands
(Pounds et al. 1996; Brandt et al. 1997) can be understood.

Further, Tanaka (1989) found
highly variable hard emission relative to the soft one in GBHC.
If the analogy holds,
the relative strength of the hard component could take any values.
This may explain a diversity in $\alpha_{\rm x}$'s of NLS1s
in the ASCA bands.

\subsection{Evolution of the disk luminosity}

It should be repeated that the disk luminosity,
increases only moderately above $L_{\rm E}$ for steadily
increasing $\dot M$.
We calculate $L$ by summing up the contribution
from each concentric annulus and plot in figure 2
as a function of $\dot m$.  Asterisks are calculated values,
while the solid curve represents the fitting formula
(see Watarai et al. 2000a for a derivation),
\begin{equation}
\label{fit}
 L({\dot m})/L_{\rm E} \simeq \left\{ \begin{array}{lcl}
    {2[1 + \ln({\dot m}/{50})]} & \mbox{for} & {\dot m} \geq 50 \\
    {\dot m}/{25} & \mbox{for} & {\dot m} < 50
\end{array} \right.
\end{equation}
The functional dependence of $L/L_{\rm E}$ on $\dot m$
is basically the same as that for GBHC with
smaller black-hole masses, thus being insensitive to $M$.

% Fig.2
\begin{figure}[t]
\hbox{\psfig{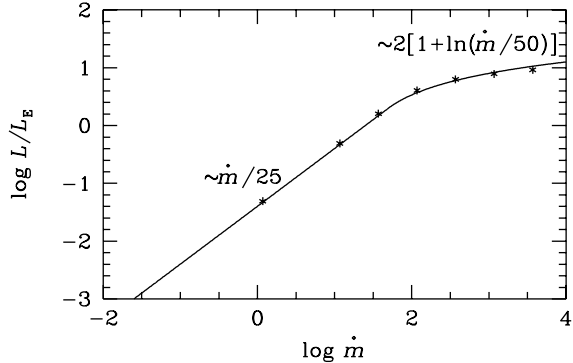}}
\caption
{
Disk luminosity as a function of $\dot m$.
Asterisks denote the calculated luminosities, whereas
the solid line shows the fitting formula [equation (\ref{fit})].
It is clear that increase in $L$ is suppressed at
$L > 2 L_{\rm E}$.
}
\end{figure}

Care must be taken when interpreting $L$, since
radiation field is likely to be anisotropic; i.e.,
radiation mostly goes out in the vertical direction to the disk plane, 
whereas matter approaches to the black hole along the disk plane.
This is a way in which super-critical accretion is possible,
but at the same time, we expect much diluted radiation going out
in the direction of the disk plane.
Blocking of the radiation by the outer disk rim provides
an additional diluting effect (Fukue 2000).
Thus, we will detect weaker flux ($< L_{\rm E}/4\pi d^2$
with $d$ being distance),
even when the total $L$ exceeds $L_{\rm E}$.
We also expect strong radiation-accelerated wind in the central region
($e.g.$, Watarai, Fukue 1999),
which will give rise to blue-shifted UV lines (Leighly 1999b).

\subsection{Fitting with the multi-color blackbody}

Now we fit the calculated disk spectra by
the extended disk blackbody model (Mineshige et al. 1994a),
assuming the following $T_{\rm eff}$ profile,
\begin{equation}
\label{teff}
  T_{\rm eff} = T_{\rm in}({r}/{r_{\rm in}})^{-p},
\end{equation}
where $r_{\rm in}$, $T_{\rm in}$, and $p$ are fitting parameters.

Note that the fitting results may slightly depend on 
the X-ray energy ranges for which fitting is made, 
but in the present study we fix the energy bands to be
either 0.1 -- 0.5 keV (for $\dot m < 10$),
0.1 -- 1.0 keV (for $10 \leq \dot m < 100$),
or 0.1 -- 3.0 keV (for $\dot m \geq 100$).

To have a good fitting result, especially in $p$,
wider spectral ranges, especially a good coverage of
soft X-ray bands, are certainly necessary.
The small black hole mass is preferable for this sort of
fitting, since then the disk temperature is roughly scaled as
$\sim M^{-1/4}$, expected from the standard-disk relation
(which is also roughly valid in the slim-disk regime).
In the fitting, $T_{\rm in}$ and $r_{\rm in}$ are basically
determined by the energy and the intensity of the
exponential roll-over, while $p$ depends on the slope 
in lower energy bands.  EUV to soft X-ray
observations are crucial for discriminating the slim disk
from the standard disk.  

We examine how the derived fitting parameters of $r_{\rm in}$,
$T_{\rm in}$, and $p$ change with an increase of a mass-flow rate
(and $L$).
The results of the fittings are summarized in table 1 and figure 3.
% Table 1
\begin{table}[b]
\small
\begin{center}
Table~1.\hspace{4pt}Results of fitting.\\
\end{center}
\vspace{6pt}
\begin{tabular*}{\columnwidth}{@{\hspace{\tabcolsep}
\extracolsep{\fill}}lccccc} 
\hline\hline\\[-6pt]
 $\dot m$ 
       & $\log T$(K)           & $T_{\rm in}$(keV) 
       & $\log r_{\rm in}$(cm) & $r_{\rm in}/r_{\rm S}$ & $p$ \\
\hline\\[-6pt]
  $10^0$     &  5.57   &  0.033  & 11.45  & 9.34 & 0.88 \\
  $10^{0.5}$ &  5.71   &  0.044  & 11.39  & 8.20 & 0.78 \\
  $10^{1.0}$ &  5.84   &  0.061  & 11.34  & 7.19 & 0.74 \\
  $10^{1.5}$ &  6.06   &  0.098  & 11.05  & 3.74 & 0.62 \\
  $10^{2.0}$ &  6.46   &  0.252  & 10.24  & 0.58 & 0.52 \\
  $10^{2.5}$ &  6.53   &  0.294  & 10.12  & 0.44 & 0.50 \\
  $10^{3.0}$ &  6.55   &  0.303  & 10.14  & 0.46 & 0.51 \\
  $10^{3.5}$ &  6.56   &  0.310  & 10.13  & 0.45 & 0.51 \\
\hline
\end{tabular*}
\end{table}

As $\dot m$ increases, so does $T_{\rm in}$ more or less gradually,
but at the same time
$r_{\rm in}$ shifts from $\sim 10 r_{\rm S}$ to $\sim 0.5 r_{\rm S}$
abruptly at $L/L_{\rm E} \sim 2$ ($\dot m \sim 50$).
We should recall that $r_{\rm in}$ stands for 
the size of the area emitting with $B_\nu(T_{\rm in})$ and not
the radius of the physical inner boundary 
(which should be always at $r_{\rm S}$).
In fact, the upper part of figure 1 has already shown that
the peak effective temperature shifts toward smaller radii
as $\dot m$ increases. 
% Fig.3
\begin{figure}[t]
\hbox{\psfig{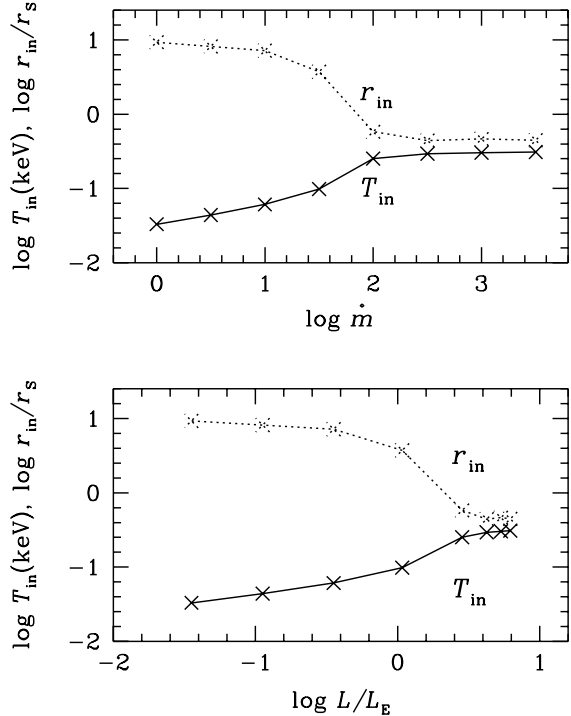}}
\caption
{
The results of the fitting, $r_{\rm in}$ (by the solid lines)
and $T_{\rm in}$ (by the dotted lines)
as functions of $\dot m$ (upper) and of $L/L_{\rm E}$, respectively.
}
\end{figure}

The effective temperature of the slim disk has a weak luminosity dependence,
$T_{\rm in} \propto M^{-1/4}\propto L^{-1/4}$,
however, $T_{\rm in}$ also has some dependence on
$\dot M/M$ at $\dot m < 50$.
We, hence, expect scatters in $T_{\rm in}$ even for the same $M$ and $L$.

\subsection{Comparison with the observations}

Hayashida (2000) analyzed the ASCA spectral data of 9 NLS1s 
and fitted them with a two spectral-component model:
a hard power-law component and a soft blackbody one.
We assume that the soft component represents the emission 
from the disk we are discussing. 
The fitting parameters of the soft component are blackbody 
temperature, $T_{\rm bb}$, 
and the size of the blackbody emitting region, $r_{\rm bb}$.
The results of the fitting are summarized in figure 4 and in table 2.  

Note that the redshift are small, $z<0.1$, for all the sources.
Among the nine sources in Hayashida (2000), we excluded
Mrk766 because of its complex spectral features and their variability
(Leighly et al. 1996).  Even for the other eight sources, it must be kept 
in mind that there may be some model dependence in the data points of 
$T_{\rm bb}$ and in $r_{\rm bb}$.  In particular,
modeling of peculiar spectral features around 1.1 keV found for some 
sources [IRAS13224, H0707, and PG1404;
see Hayashida (2000) and reference therein]
affect the parameter $T_{\rm bb}$ significantly, {\it e.g.},
$T_{\rm bb}$ for H0707 might be about 30\% higher than the value 
listed in table 2, if we model the feature as absorption edges.  
On the other hand, sources of which soft component 
are not so prominent, $e.g.$, IZw1, the results will be affected,
if we allow
some deviation from the perfect power law model for the hard component.
Calibration problem of ASCA SIS may add extra error on the results, too.  
These points would be addressed in a separate paper. 

  The theoretical expectations based on the slim disk model
are also plotted by the solid and dashed lines in figure 4.
Since inclination angles are not known, we assume
the inclination angle of $i = 60^\circ$ in the theoretical models;
i.e., we set $r_{\rm bb} = r_{\rm in}\sqrt{\cos i}$.
Likewise, we set $T_{\rm bb}= T_{\rm in}$, although
the former may be systematically higher than the latter,
$T_{\rm in} \simeq (1.1-1.3) T_{\rm bb}$.  Accordingly,
the value of $r_{\rm bb}$ should be smaller, for
$L \propto r_{\rm bb}^2T_{\rm bb}^4$ is kept roughly constant.
It might also be noted that the combination of $r_{\rm bb}T_{\rm bb}^4$,
which is proportional to $L/L_{\rm E}$, is mass-independent,
since $r_{\rm bb}$ ($\propto r_{\rm in}$) can be scaled with
$r_{\rm S}\propto M$ and $T_{\rm bb}\sim T_{\rm in}\propto M^{-1/4}$.
Therefore, constant $\dot m$ lines give a
relation, $T_{\rm bb} \propto r_{\rm bb}^{-1/4}$.

On the theoretical grounds, various kinds of GR effects
and Compton scattering should affect the results. 
When scattering opacity dominates over absorption,
the emergent spectra deviate from blackbody one (see Rybicki, Lightman 1979).
In a hot disk with Compton $y > 1$, moreover, 
inverse Compton scattering should modify the disk spectra
in the high-energy tail,
thereby yielding, at most, a factor of 2 $increase$ in $T_{\rm bb}$,
while the total flux ($\propto r_{\rm bb}^2 T_{\rm bb}^4$) is kept constant;
i.e., a factor of 4 $decrease$ in $r_{\rm bb}$
(Ross et al. 1992; Shimura, Takahara 1995).

GR effects (such as gravitational redshift,
Doppler boosting, gravitational focusing due to ray bending)
produce complex effects.  We calculate a slim disk model
with face-on geometry, for which only gravitational redshift works,
finding systematically $larger$ $r_{\rm bb}$ ($\sim 3 r_{\rm S}$) 
by a factor of $\sim 6$
and thus $lower$ $T_{\rm bb}$ ($\sim 0.12$ keV) 
by a factor of $\sim 2.5$ even for $\dot m \gsim 100$.
In other words, most of radiation from inside $3 r_{\rm S}$ vanishes 
for this case.
[Although photon energy gets lost partly, the total flux does not drop much
because of flatter $T_{\rm eff}$ profile, $T_{\rm eff}(r)\propto r^{-1/2}$.]
Since Compton and GR effects work 
independently in opposite ways for a face-on disk, these effects cause 
decrease in $T_{\rm bb}$ (by a factor of 2.5/2)
and increase in $r_{\rm bb}$.  
Corrections to remove Compton and GR effects from the data points are, hence,
$\Delta \log T_{\rm bb} \lsim +0.1$ and 
$\Delta \log r_{\rm bb} \lsim -0.2$,
which is indicated by the lower-right-ward arrow in figure 4.
(Since a model predicts lower $T_{\rm bb}$ by these effects,
we need higher $\dot m$ to reproduce the observed $T_{\rm bb}$.)

For disks with non-zero inclination angles around a Schwarzschild black hole, 
Doppler boosting enhances
radiation, yielding, at most, a factor of $\sim 1.4$ $increase$ 
in $T_{\rm bb}$ compared with face-on disks,
while the total flux kept roughly constant (Sun, Malkan 1989).
We thus expect slight $increase$ in $T_{\rm bb}$
and $decrease$ in $r_{\rm bb}$ for a fixed $r_{\rm bb}^2T_{\rm bb}^4$
compared with face-on disks.  Required corrections are
$\Delta\log T_{\rm bb} \lsim -0.2$ and 
$\Delta\log r_{\rm bb} \lsim +0.4$, which 
is indicated by the upper-left-ward arrow in figure 4.

% Table 2
\begin{table}[b]
\small
\begin{center}
Table~2.\hspace{4pt}Results of ASCA Observations.\\
\end{center}
\vspace{6pt}
\begin{tabular*}{\columnwidth}{@{\hspace{\tabcolsep}
\extracolsep{\fill}}lcccc} 
\hline\hline\\[-6pt]
 source/$\dot m$ & $T_{\rm bb}$(keV) & $r_{\rm bb}$(cm) 
                            & $\log(T_{\rm bb}^4r_{\rm bb})$ \\
\hline\\[-6pt]
 IZw1     & 0.100 & $9.75\times 10^{10}$ & 6.99 \\
 REJ1034  & 0.107 & $1.04\times 10^{11}$ & 7.14 \\
 PG1440   & 0.097 & $2.08\times 10^{11}$ & 7.28 \\
IRAS13224 & 0.098 & $2.55\times 10^{11}$ & 7.38 \\
 H0707    & 0.091 & $3.97\times 10^{11}$ & 7.44 \\
 PG1211   & 0.109 & $2.21\times 10^{11}$ & 7.50 \\
 PG1244   & 0.191 & $2.39\times 10^{10}$ & 7.51 \\
 PG1404   & 0.081 & $9.50\times 10^{11}$ & 7.62 \\
\hline
\end{tabular*}
\end{table}
% Fig.4
\begin{figure}[t]
\hbox{\psfig{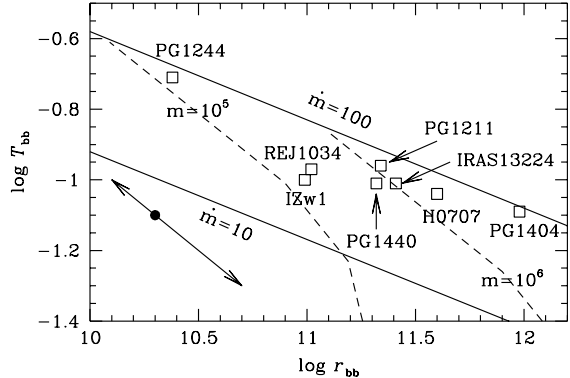}}
\caption
{
The $r_{\rm bb}$-$T_{\rm bb}$ diagram of the observational
data of NLS1s.  The loci of $\dot m=10$ and 100 (by the solid lines)
and those of $m (\equiv M/M_\odot)= 10^5$ and $10^6M_\odot$ 
(by the dotted lines)
obtained by the slim disk are also shown ($i=60^\circ$ is assumed).
The arrow headed for upper-left (or lower-right)
from the filled circle in the lower-left corner indicates
the direction of the correction for that point to remove Compton and GR effects
in the case of a face-on (nearly edge-on) disk.
}
\end{figure}

It is evident in figure 4 that all NLS1s have relatively 
large $\dot m$.  In other words,
NLS1s shine at luminosities close to the Eddington (Hayashida 2000).
Especially, PG1404, H0707, and IRAS13224,
all having large $\dot m$ in figure 4
and enhanced soft components,
are known to exhibit giant amplitude variability.
In fact, Hayashida (2000) found that
the mass estimated by the variability
(Hayashida et al. 1998) significantly deviate from that
estimated by equating $r_{\rm bb} = 0.5 r_{\rm S}$ in those sources,
suggesting anomalously large variability amplitudes 
compared with other NLS1s with a similar black-hole mass.
This tendency indicates a correlation between $L/L_{\rm E}$
and the variability amplitudes (and soft X-ray excess over hard X-rays,
see below).

Despite ambiguities in data points in figure 4, we may conclude 
that all (or, at least, most of) the NLS1s fall on
the region with $\dot m > 10$,
supporting our view that slim disks emit substantial fraction of
soft X-rays at $\sim L_{\rm E}$ in NLS1s.
The observed steep slopes, $a_{\rm os}$ (see section 3.2), rather enhances
our conclusion of high $\dot m$, since it means more optical radiation 
and thus higher $\dot m$ than that expected solely from X-rays.

\section{Variability of NLS1s}

The strength of the soft excess is correlated with the variability
amplitudes; objects with strong soft excesses show higher amplitude
variability (Leighly 1999b).
Hence, successful models for NLS1s should account for the presence of 
soft X-ray variability.
Leighly (1999a) summarize the ASCA observations of NLS1s, 
finding systematically large excess variance in NLS1s.
This fact can be, in principle, explained in terms of small black-hole masses,
however, origin of giant-amplitude variations observed in some NLS1s,
including IRAS 13224-3809 (Boller et al. 1997, see also Otani et al. 1996)
and PHL 1092 (Brandt et al. 1999),
seems to require additional explanation (see section 3.5).
The lack of optical variability in IRAS 13224-3809
poses a problem (Young et at. 1999),
although both of the soft and hard components (in the ASCA band) are known to
vary nearly simultaneously in the same source (Otani et al. 1996).
Relativistic boosting of otherwise modest variations may
produce extreme variability of this source (Boller et al. 1997).

The light curves of NLS1s, especially those exhibiting extreme variations,
 have the following characteristics:
\par
\noindent{\bf (1)}
The light curves seem to be composed of numerous flares (or shots).
\par
\noindent{\bf (2)}
The flares do not have an identical time profile but
show a variety of flare amplitudes and durations.
\par
\noindent{\bf (3)}
Temporal distribution of each flare is apparently random.
\par

Interestingly, these features are the same as those
found in GBHC during the low (hard) and very high states
(see, $e.g.$, van der Klis 1995).
Fluctuations in the soft X-ray component (of a disk origin) is observed 
only in the very high state, 
while the low-state disk has no soft component.
In this sense, NLS1s are more like GBHC during the very high state.

To reproduce theses characteristics of the light curves,
Kawaguchi et al. (2000) proposed that
fractal magnetic-field structure spontaneously arises in
an optically thin, advection-dominated flow and
produces sporadic magnetic flare events, like solar flares, 
via magnetic reconnection.  In fact,
dynamics of magnetic fields in ADAFs is known to be well represented by
a cellular-automaton model based on the notion of the
Self-Organized Criticality
(Mineshige et al. 1994b, Kawaguchi et al. 1998).
It is of great importance to note that
both of optically thick and thin ADAFs may share common
time-dependent properties, since they are both
dynamical systems in the sense that radial accretion velocity,
$v_r$, is comparable to the free-fall velocity and sound speed,
as long as $\alpha$ is not small.
We, here, argue that the same MHD model can be relevant to the
slim disk, thus explaining the variability of NLS1s, as well.

If fluctuations are of magnetic origin,
large-amplitude fluctuations indicate relatively large 
field energy compared to the radiation energy.
Suppose that gas cloud of a unit mass falls onto a black hole.
Since radiative cooling is efficient in the standard-type disk,
energy that is emitted away 
is comparable to the release of the gravitational energy, 
$E_{\rm rad} \sim E_{\rm grav} (\sim GM/r)$.  
As a result, the internal energy of gas should be much lower than 
the gravitational energy, $E_{\rm gas} \ll E_{\rm grav}$.  
Since magnetic field energy
will be, at most, the same order of the gas energy,
$E_{\rm mag}\ltsim E_{\rm gas}$, we have 
$E_{\rm mag} \ll E_{\rm grav} \sim E_{\rm rad}$.
If fluctuations occur due to sporadic release of magnetic energy 
via magnetic reconnection,
this inequality explains small fluctuations in the standard disk.

In the ADAF state, both optically thick and thin ones, 
radiative cooling is inefficient.  
Suppose again that gas cloud of a unit mass falls onto a black hole.
A distinction exists in what carries energy to a hole;
it is internal energy of gas for the case of the optically thin ADAF, 
whereas it is trapped photons for the slim disks.  
In any cases, internal energy of gas or of trapped photons
is of the order of gravitational energy which
much exceeds radiation energy which is emitted away, 
$(E_{\rm rad})_{\rm trapped} > (E_{\rm rad})_{\rm out}$.
Since again $E_{\rm mag}\ltsim (E_{\rm rad})_{\rm trapped}$, 
moderately large magnetic energy compared with the energy radiated away
is expected, $E_{\rm mag} \gtsim (E_{\rm rad})_{\rm out}$.
The larger $\dot m$ is, the more becomes the ratio of
$(E_{\rm rad})_{\rm trapped} / (E_{\rm rad})_{\rm out}$.
Namely, large fluctuations in soft X-ray emission are inevitable at
high $\dot m$, consistent with the observations of NLS1s.

In contrast, the standard-type disk (which may exist in BLS1s) is 
stable and, hence, is not variable.
Fluctuations from BLS1s are thus due to variation of
incident power-law radiation
from disk coronae, which are presumably filled with magnetic fields and
thus variable for the same reason mentioned above.  
In terms of variability,
therefore, the disk main body is passive in BLS1s while it is active
in NLS1s. To elucidate the theory distinguishing these two, however,
realistic modeling of the disk-corona structure is needed as future work.

\section{Outstanding issues}

%Table 3
\begin{table*}[t]
\small
\begin{center}
Table~3.\hspace{4pt}Basic disk models and their properties.\\
\end{center}
\vspace{6pt}
\begin{tabular*}{\textwidth}{@{\hspace{\tabcolsep}
\extracolsep{\fill}}llll} 
\hline\hline\\[-6pt]
 disk models    &  standard disk ($\tau>1$) 
                & ADAF ($\tau<1$) & ADAF ($\tau>1)$ \\
 \hline
 energy balance & $Q_{\rm vis}=Q_{\rm rad} \gg Q_{\rm adv}$ 
                & $Q_{\rm vis}=Q_{\rm adv} \gg Q_{\rm rad}$ 
                & $Q_{\rm vis}=Q_{\rm adv} \gg Q_{\rm rad}$ \\
 disk height    &  $H \ll r$           & $H \lsim r$     &  $H \lsim r$ \\
 electron temperature % $T_{\rm elec}$(K)
                & $\sim 10^6M_5^{-\frac 14}
                                             (r/r_{\rm S})^{-\frac 34}$K
                & $\lsim 10^{10}$K
		& $\sim 2\times 10^6M_5^{-\frac 14}
                                            (r/r_{\rm S})^{-\frac 12}$K \\
 ion temperature      % $T_{\rm ion}$(K)  
                & $\sim 10^6M_5^{-\frac 14}
                                             (r/r_{\rm S})^{-\frac 34}$K
                & $\sim 10^{12} (r/r_{\rm S})^{-1}$K
		& $\sim 2\times 10^6M_5^{-\frac 14}
                                            (r/r_{\rm S})^{-\frac 12}$K \\
 luminosity     & $L \propto {\dot M}$ 
                & $L \propto ({\dot M})^2$
	        & $L \propto \ln{\dot M}$ \\
 spectrum       &  blackbody       & power-law        & blackbody     \\
final energy form  & emitted radiation
                   & internal energy of gas
                   & trapped radiation \\
 magnetic energy & $E_{\rm mag} \ll E_{\rm grav} \sim E_{\rm rad}$    
                 & $E_{\rm mag} \sim E_{\rm grav} \gg  E_{\rm rad}$ 
                 & $E_{\rm mag} \sim  E_{\rm grav}  >   E_{\rm rad}$ \\
fluctuation     &  small           &   large          &  large    \\ 
\hline
\end{tabular*}
\end{table*}

\subsection{Radiation transfer}
In a hot and dense layer with a very large Thomson depth ($\tau$),
there are several important effects to be considered in addition
to Compton scattering effects.
First, the photon diffusion time ($\sim H\tau/c$) may exceed the
accretion time.  
\begin{equation}
  \tau_{\rm dif} \simeq 4\times 10^4 \Sigma_4 (H/r)(r/10 r_{\rm S})M_5{\rm s}
\end{equation}
and
\begin{equation}
  \tau_{\rm acc} \simeq 5\times 10^2 (v_r/0.1 v_{\rm ff})^{-1}
                                       (r/10 r_{\rm S})^{3/2}M_5{\rm s},
\end{equation}
where $\Sigma_4 \equiv \Sigma/10^4$g cm$^{-2}$
and $v_r$ is the accretion (radial) velocity.
Then, the amount of photons which can be emitted
to space is limited so that the disk luminosity gets even lower, say
below $L_{\rm E}$, even for large $\dot m$.
Such effects as those arising due to a finite diffusion time are omitted
in the present study 
but this treatment is reasonable if efficient
convective mixing of gas and radiation occurs so that photons generated
on the equatorial plane can reach the disk surface on dynamical timescales.
Such effects should be examined in future, however.

Second,
thermalization timescales may exceed the accretion time
for $\alpha \gsim 0.03$ (Beloborodov 1998).
Then, disk temperature gets remarkably high, $\sim 10^8$K
even for the parameters of AGN, producing much higher-energy radiation.
However, such high-energy emission spectra do not reconcile with 
the observed soft spectra of NLS1s with soft excess of $kT \sim 100$eV.
Probably, again convective mixing of gas and radiation 
promotes rapid thermalization between them.
Or alternatively, the viscosity parameter could be small,
$\alpha < 0.03$, for which thermalization occurs within accretion time.

Finally, radiation field is more likely to be anisotropic; i.e.,
radiation mostly goes out in the vertical direction to the disk plane, 
whereas matter approaches to the black hole along the disk plane.
This much reduces flux to an observer viewed from the direction with
large inclination angles.  Sub-Eddington luminosity is expected
even when the total $L$ exceeds $L_{\rm E}$.

\subsection{Limit-cycle oscillations}
Since the standard and slim-disk branches form an $\bf S$-shape
in the thermal equilibrium curves (Abramowicz et al. 1988),
there is a possibility of limit-cycle oscillations between
the two branches (Honma et al. 1991; Szuszkiewicz, Miller 1998).
Such oscillations, if occurs, will possibly explain quasi-periodic
outburst behavior, which may explain transient AGN phenomena
in otherwise normal (inactive) galactic nuclei
(see, $e.g.$, Komossa, Bade 1999; Komossa, Greiner 1999),
although repeated outbursts will produce permanent AGN-typical
NLR emission lines, which are not observed in some sources.
Alternatively, the $\bf S$-shaped curve may produce soft-hard
transitions; i.e., transitions from the slim disk to the
optically thin ADAF (Takeuchi, Mineshige 1998).
These possibilities should be elucidated in future.

\subsection{Implications on the cosmological evolution}
It is often argued that for a BH to increase its mass
by a factor of $\sim e$
it takes $t_{\rm E} \equiv M/{\dot M_{\rm E}} \sim 10^8$yr,
since the maximum accretion rate onto a BH is regulated by
the Eddington rate, ${\dot M_{\rm E}}\sim 10 L_{\rm E}/c^2$.
Then, it takes quite a long time,
$\sim 2\times 10^9$yr, to make a $10^9M_\odot$ BH 
out of $\sim 10M_\odot$ BH.
Much shorter time is enough, however,
if supercritical accretion is allowed.
Indeed, supercritical accretion is unavoidable
to account for the present quasar population
(e.g. Kauffmann, Haehnelt 1999).
It might be that there used to be numerous NLS1s in the past.
The observational consequences need to be investigated.

\section{Conclusions}

To summarize, we discussed various aspects of the slim disk
in the context of NLS1s in comparison with the other
disk models (see table 3). 
The followings are the main conclusions.
\vspace{-6pt}

\begin{enumerate}
\item
The disk begins to exhibit the slim-disk properties, when
the luminosity moderately exceeds $L_{\rm E}$.  
Then, substantial amount of radiation is expected to originate
from inside the marginally
stable last circular orbit at $3r_{\rm S}$.  Therefore,
a small $r_{\rm bb} < 3r_{\rm S}$ does
not necessarily mean the presence of a Kerr hole.

\item
If we fit the spectrum of the slim disk
with the disk blackbody model, the derived size of the
X-ray emitting region will be $r_{\rm in} \lsim 3 r_{\rm S}$
and the maximum temperature will be
$kT_{\rm in} \sim 0.2M_5^{-1/4}$keV.
Further, $p$ $(\equiv d\ln T_{\rm eff}/d\ln r)$ decreases from 
$0.75$ to $\sim 0.5$ as $\dot M$ increases.

\item
All the ASCA data of NLS1s fall on the region with
supercritical accretion, $\dot m > 10$,
supporting that the disks in NLS1s are likely to be slim disks.

\item
The slim disk can also produce large fluctuations 
because release of magnetic-field energy (which will produce
fluctuations via magnetic flares) dominates over
the energy release due to persistent emission.

\end{enumerate}

\vspace{1pc} \par
We are grateful for A. Laor and an anonymous referee for
useful comments.
This work was supported in part by the Grants-in Aid of the
Ministry of Education, Science, Sports, and Culture of Japan
(10640228, SM).

\section*{References}
\small

\re
Abramowicz M.A., Czerny B., Losota J.P., Szuszkiewicz E. 1988,
   ApJ 332, 646

\re
Beloborodov A.M. 1998, MNRAS 297, 739

\re
Boller Th., Brandt W.N., Fabian A.C., Fink H.H. 1997,
  MNRAS 289, 393

\re
Boller Th., Brandt W.N., Fink H.H. 1996, A\&A 305, 53

\re
Boroson T.A., Green R.F. 1992, ApJS 80, 109

\re
Brandt W.N. in High-Energy Processes in Accreting Black Holes
  (ASP Conf. Ser. 161) ed J. Poutanen, R. Svensson (ASP, San Francisco) p166

\re
Brandt W.N., Boller Th., 1998, Astron. Nachr. 319, 163

\re
Brandt W.N., Boller Th., Fabian A.C., Ruszkowski 1999, MNRAS 303, L53

\re
Brandt W.N., Mathur S., Elvis M. 1997, MNRAS 285, L25

\re
Brandt et al. 2000, ApJ 528, 637

\re
Czerny B., Elvis M. 1987, ApJ 321, 305

\re
Fukue J. 2000, PASJ submitted

\re
Grupe G.D., Beuermann K., Thomas H.-C., Mannheim K. Fink H.H. 
 1998, A\&A 330, 25

\re
Grupe G.D., Beuermann K., Mannheim K., Thomas H.-C. 1999, 
  A\&A 350, 805 

\re
Hayashida K. 2000, in Proc. of 32nd COSPAR Meeting:
 Broad Band X-Ray Spectra of Cosmic Sources,
 ed K. Makishima (Adv. Space Research), 25, 489

\re
Hayashida K. et al. 1998, ApJ 500, 642

\re
Halpern J.P., Oke J.B. 1987, ApJ 312, 91

\re
Honma F., Matsumoto R., Kato S. 1991, PASJ 43, 147

\re
Kato S., Fukue J., Mineshige S. 1998,
Black-Hole Accretion Disks (Kyoto University Press, Kyoto)

\re
Kaspi S., Smith P.S., Maoz D., Netzer H., Jannuzi B.T. 1996,
  ApJ 471, 75

\re
Kauffmann G., Haehnelt M. 1999, (astro-ph/9906493)

\re
Kawaguchi T., Mineshige S., Machida M., Matsumoto R., Shibata K.
  2000, PASJ in press

\re
Kawaguchi T., Mineshige S., Umemura M., Turner E.L.  1998, ApJ 504, 671

\re
Komossa S., Bade N. 1999, A\&A in press

\re
Komossa S., Greiner J. 1999, A\&A 349, L45

\re
Laor A. 1998 in astro-ph/9810227

\re
Laor A., Fiore F., Elvis M., Wilkes B.J., McDowell J.C. 1997,
 ApJ 477, 93

\re
Leighly K.M. 1999a, ApJS accepted (astro-ph/9907294)

\re
Leighly K.M. 1999b, ApJS accepted (astro-ph/9907295)

\re
Leighly K.M., Mushotzky R.F., Yaqoob T., Kunieda H., Edelson R.
  1996, ApJ, 469, L147.

\re
Matsumoto R., Kato S., Fukue J., Okazaki A.T. 1984, PASJ 36, 71

\re
Mineshige S., Hirano A., Kitamoto S.,
    Yamada T.T., Fukue J. 1994a, ApJ 426, 308

\re
Mineshige S., Kusunose M., Matsumoto R. 1995, ApJ 445, L43

\re
Mineshige S., Takeuchi M., Nishimori H. 1994b, ApJ 435, L125

\re
Narayan R., Yi I. 1995, ApJ 444, 231

\re
Osterbrock D.E., Pogge R.W. 1985, ApJ 297, 166

\re
Otani C., Kii T., Miya K. 1996
 in R\"ontgenstrahlung from the Universe
 (MPE Report 263), ed H.U. Zimmermann, J.E. Tr\"umper, H. Yorke
  (MPE Press, Garching) p491

\re
Paczy\'{n}ski B., Wiita P.J. 1980, A\&A 88, 23

\re
Pounds K.A., Done C., Osborne J. 1996, MNRAS 277, L5

\re
Ross R., Fabian A.C., Mineshige S. 1992, MNRAS 258, 189

\re
Rybicki G.B., Lightman A.P. 1979,
 Radiative Processes in Astrophysics (John Wiley \& Sons, New York)

\re
Shakura N.I., Sunyaev R.A. 1973, A\&A 24, 337

\re
Shimura T., Takahara F. 1995, ApJ 440, 610

\re
Sun W.-H., Malkan M.A. 1989, ApJ 346, 68

\re
Szuszkiewicz E., Malkan M.A., Abramowicz M.A. 1996, ApJ 458, 474

\re
Szuszkiewicz E., Miller J.C. 1998, MNRAS 298, 888

\re
Takeuchi M., Mineshige S. 1998, ApJ 505, L19

\re
Tanaka Y.  1989, in Proc. 23rd ESLAB Symp. on Two Topics in
  X-Ray Astronomy, ed J. Hunt, B. Battrick (ESA SP-296, Dordrecht) p3

\re
van der Klis M. 1995, in X-Ray Binaries, 
   ed W.H.G. Lewin et al. (Cambridge University Press, Cambridge) p58 

\re
Wang J., Szuszkiewicz E. et al. 1999, ApJ 522, 839

\re
Watarai K., Fukue J. 1999, PASJ 51, 725

\re
Watarai K., Takeuchi M., Fukue J., Mineshige, S. 2000a, PASJ 52, in press

\re
Watarai K. et al. 2000b, in preparation

\re
Wills B., Laor A., Brotherton M.S., Wills D., Wikes B.J.,
  Ferland G.J., Shang Z. 1999, ApJ 515, L53

\re
Young A.J., Crawford C.S., Fabian A.C., Brandt W.N., O'Brien P.T.
 1999, MNRAS 304, L46

\end{document}